\documentclass[a4paper]{jpconf}
\usepackage{xcolor}
\usepackage{graphicx}
\usepackage{bm}
\usepackage{siunitx}
\pagecolor{white}
\usepackage{url}
\usepackage{hyperref}

\begin{document}
\title{LASY: LAser manipulations made eaSY}

\author{M Thévenet$^{1}$, I A Andriyash$^{2}$, L Fedeli$^{3}$, A Ferran Pousa$^{1}$, A Huebl$^{4}$, S Jalas$^{1}$, M Kirchen$^{1}$, R Lehe$^{4}$, R J Shalloo$^{1}$, A Sinn$^{1,5}$ and J-L Vay$^{4}$}

\address{$^{1}$Deutsches Elektronen-Synchrotron DESY, Notkestr. 85, 22607 Hamburg, Germany}
\address{$^{2}$Laboratoire d’Optique Appliquée, ENSTA Paris, CNRS, Ecole polytechnique, Institut Polytechnique de Paris, 828 Bd des Maréchaux, 91762 Palaiseau, France}
\address{$^{3}$Université Paris-Saclay, CEA, LIDYL, 91191 Gif-sur-Yvette, France}
\address{$^{4}$Lawrence Berkeley National Laboratory LBNL, Berkeley, California 94720, USA}
\address{$^{5}$Universität Hamburg UHH, Mittelweg 177, 20148 Hamburg, Germany}

\ead{maxence.thevenet@desy.de}

\begin{abstract}

Using realistic laser profiles for simulations of laser-plasma interaction is critical to reproduce experimental measurements, but the interface between experiments and simulations can be challenging. Similarly, start-to-end simulations with different codes may require error-prone manipulations to convert between different representations of a laser pulse. In this work, we propose LASY, an open-source Python library to simplify these workflows. Developed through an international collaboration between experimental, theoretical and computational physicists, LASY can be used to initialize a laser profile from an experimental measurement, from a simulation, or from analytics, manipulate it, and write it into a file in compliance with the openPMD standard. This profile can then be used as an input of a simulation code.
\end{abstract}

\section{Introduction}

Rapid progress in laser technology\,\cite{strickland1985,danson2019} has powered advances in laser-plasma interaction research, in particular towards inertial confinement fusion\,\cite{abu2022}, laboratory astrophysics\,\cite{dong2012,takabe2021} and plasma acceleration\,\cite{macchi2013,gonsalves2019}. While cutting-edge laser systems typically demonstrate non-ideal spatial and temporal profiles, most numerical simulations use analytic profiles (e.g. Gaussian), resulting in sub-optimal accuracy\,\cite{beaurepaire2015,ferri2016}. This is partially because the use of realistic laser profiles requires operations (construct a profiles from measurements, propagate to desired location, remove noise, etc.) that can be time-consuming and error-prone. Similarly, for start-to-end simulations, propagating a laser pulse through various codes using different laser models requires specific manipulations (convert from electric field to vector potential, from full field to an envelope model\,\cite{benedetti2017}, employ different file formats, etc.). The LASY library --standing for \emph{LAser manipulations made eaSY}--, proposes a user-friendly implementation of these manipulations, to facilitate realistic and start-to-end simulations.

Internally in LASY, the laser pulse is represented by its complex envelope $\mathcal{E}$, related to the laser electric field by:
\begin{equation}
\boldsymbol{E_{\perp}}(x,y,t) = \mathrm{Re}\left[ \mathcal{E}(x,y,t) e^{-i\omega_0t}\boldsymbol{p}\right]
\end{equation}
where $\boldsymbol{E_{\perp}}$ is the transverse electric field of the laser pulse, $\omega_0 = 2\pi c/\lambda_0$ is the reference angular frequency (typically close to the laser central frequency), $\lambda_0$ is the reference laser wavelength, and $\boldsymbol{p}=(p_x,p_y)$ is the polarization vector. This vector is complex and normalized ($|\boldsymbol{p}|=1$), and allows for an arbitrary polarization state: for any $\psi$, $\boldsymbol{p}=(e^{i\psi},0)$ gives linear polarization in x,  $(\frac{e^{i\psi}}{\sqrt{2}},\frac{e^{i(\psi+\pi/2)}}{\sqrt{2}})$ gives circular polarization, etc. ($\psi$ controls the carrier-envelope phase).

\begin{figure}[h]
\begin{center}
\includegraphics[width=38pc]{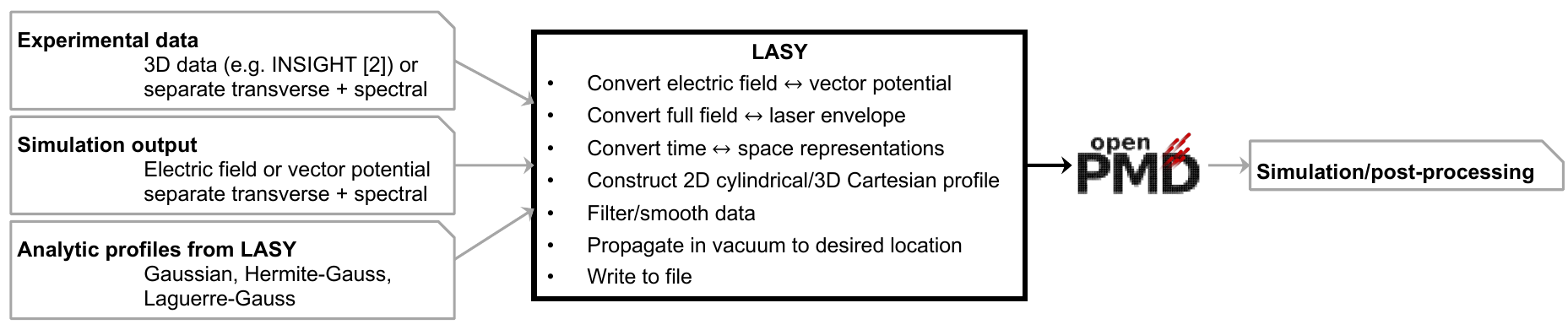}
\caption{Schematic of the LASY structure: laser profiles from various sources, e.g., generated by LASY or from a measurement, are used as input to the library. LASY can then perform a number of operations on the profile before writing it to a standardized format.}
\end{center}\label{fig:structure}
\end{figure}

The general structure and capabilities of LASY are shown in Fig.\,\ref{fig:structure}. The user specifies the physical properties of a laser pulse when constructing an instance of class \verb|Profile| and its derived classes --supporting experimental profiles, simulation results and a collection of analytical profiles. This instance is used to build a \verb|Laser| instance, the main class in LASY, which performs most operations. The propagation in vacuum is powered by the open-source Axiprop library\,\cite{oubrerie2022}. Common operations are also available as stand-alone \verb|utils| through a lightweight interface. A method of the \verb|Laser| class allows writing to file in the openPMD standard\,\cite{openPMDstandard}.

Building on the base standard of openPMD, the \verb|LaserEnvelope| extension\,\cite{openpmdLaser} provides additional metadata to encode the polarization vector, the central frequency of the laser pulse, the representation, etc. The standard supports representing the laser pulse in electric field or normalized vector potential, in $(x,y,t)$ or $(x,y,z)$ space. Simulation codes that already support reading such openPMD files are currently FBPIC\,\cite{lehe2016}, WarpX\,\cite{myers2021}, HiPACE++\,\cite{diederichs2022}, and Wake-T\,\cite{pousa2019}. Support was added to openPMD tools and visualization frameworks.

In what follows, the current capabilities of LASY are demonstrated on two common workflows: using an experimentally measured profile in a particle-in-cell (PIC) simulation\,\cite{birdsall2018}, and transferring a laser pulse from an electromagnetic representation (full electric field, including oscillations at the carrier frequency) to a quasi-static representation (envelope of the vector potential).

\section{From measurement to simulation}

\begin{figure}[h]
\begin{center}
\includegraphics[width=38pc]{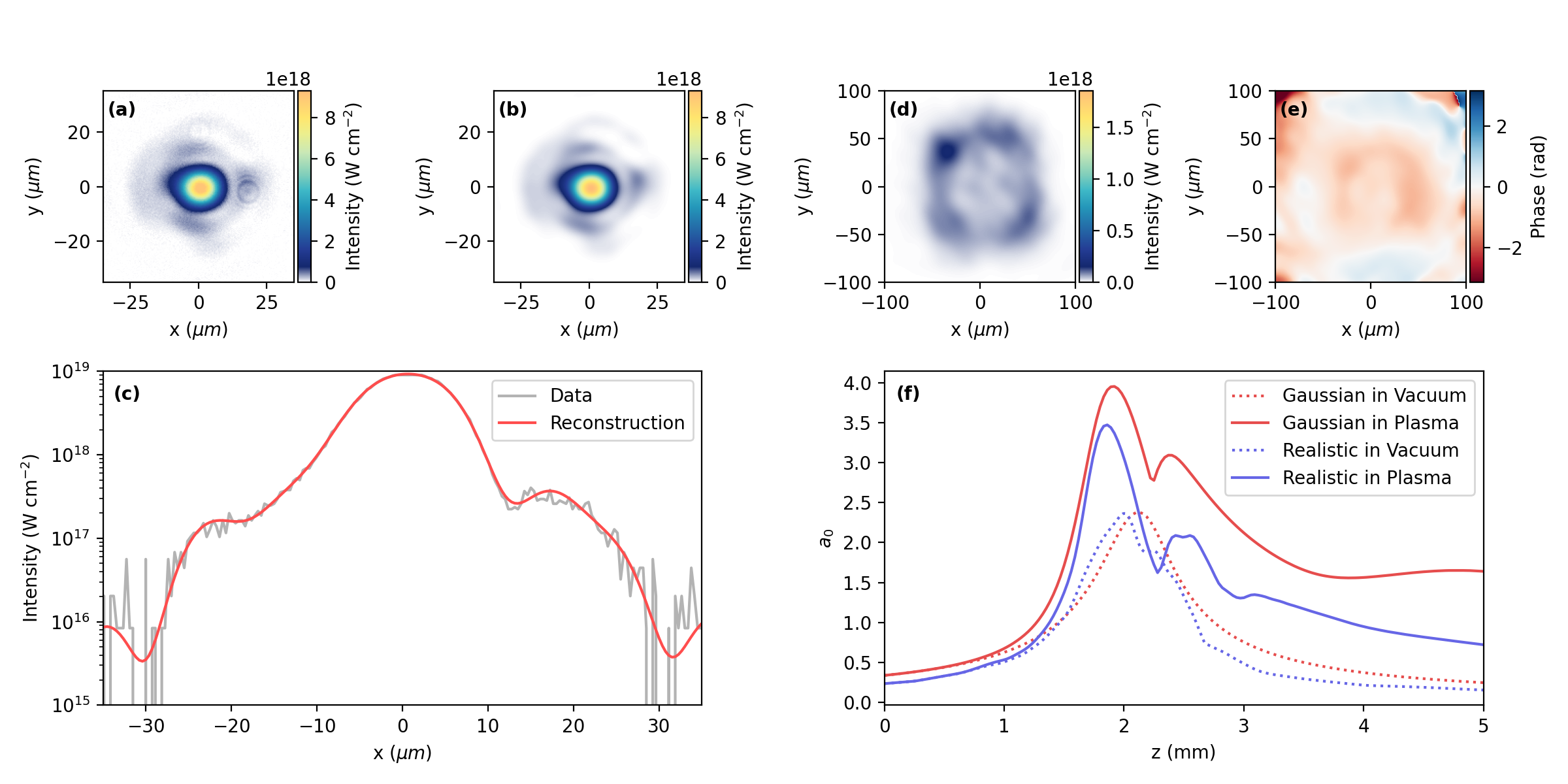}
\caption{Demonstration of the experimental data cleaning built into LASY. (a) the background subtracted laser intensity profile, calculated from a measured fluence profile using a focal spot camera (b) the pulse reconstructed from a set of Hermite-Gaussian modes showing a reduction in noise and an improved smoothness. (c) shows a lineout of the intensity profile along the x-axis on a linear and logarithmic y-scale. (d) and (e) show the measured intensity and phase of the profile after a propagation of 2~mm upstream of focus (the associated quadratic phase was removed). The propagation of this measured profile in a plasma and in vacuum are shown in (f), and compared with an equivalent Gaussian profile.}
\end{center}\label{fig:experimental}
\end{figure}


In this section, the spatial and temporal profiles of a \qty{25}{\tera\watt} Titanium Sapphire laser system are inserted into a simulation with the help of LASY, see Fig.~\ref{fig:experimental}. The laser has a central wavelength of \qty{805}{\nano\meter} and produces up to \qty{500}{\milli\joule} of energy compressed to a pulse of duration $\sim$ \qty{27}{\femto\second} ~\cite{Bohlen2022}. The pulse has a super-Gaussian near-field profile of diameter \qty{40}{\milli\meter} and is focused in the vacuum chamber using an off-axis paraboloid of focal length \qty{500}{\milli\meter} to provide a spot with a $1/e^2$ intensity radius of $\sim$ \qty{9}{\micro\meter}. 

Images of the focal intensity profile are recorded using a focal spot camera comprising a CCD sensor onto which the focal spot intensity profile is re-imaged using a $10 \times$ infinity corrected microscope objective paired with a \qty{200}{\milli\meter} tube lens. The structure is mounted onto a 3-axis motorized stage to track the focus over several millimeters. The spatial resolution of the camera is calibrated using a USAF test chart. The background-subtracted fluence images from the focal spot camera can be directly imported into python as \verb|numpy| arrays and then added to LASY as individual \emph{transverse profiles}, see  Fig.~\ref{fig:experimental} (a). An additional step can be undertaken at this stage to reduce the noise of the measured profile (due to e.g. the sensor's dynamic range) using a Hermite-Gaussian modal decomposition. This is performed by calculating a user-defined number of Hermite-Gaussian modes and projecting these onto the measured fluence profile to extract the contribution of each mode to the final profile ~\cite{dickson2022}. The full transverse profile may then be reconstructed by a weighted sum of these modes, with the result of this reconstruction shown in Fig.~\ref{fig:experimental} (b). Figure~\ref{fig:experimental} (c) shows a lineout of the data and the reconstructed profile demonstrating the improvement in the smoothness of the signal and an increase in the dynamic range of the array. Both of these improvements are important for reducing numerical noise later in the simulation. 

While the method above reproduces only the spatial amplitude, the representation of real laser pulses in simulations can be greatly improved using a phase retrieval algorithm such as the Gerchberg-Saxton algorithm\,\cite{gerchberg1972}. Several measured laser fluence profiles can then be combined and used to reconstruct the spatial phase profile of the pulse. A version of the Gerchberg-Saxton algorithm based on a Hermite-Gaussian modal decomposition, GSA-MD, has recently been proposed ~\cite{Moulanier2023} and is currently being implemented into LASY. The method incorporates an analytic description of the propagation of the Hermite-Gaussian modes. Figure~\ref{fig:experimental} (d) and (e) show the spatial intensity and phase of the laser reconstructed $\sim$\,\qty{2}{\milli\meter} upstream from focus using a series of five focal scan images and the GSA-MD algorithm. The quadratic phase (dominating because the pulse is out of focus) was removed to better elucidate the phase aberrations present in the pulse. This measured transverse profile was combined with a Gaussian temporal profile within LASY. The resulting 3D array was written to a file in compliance with the openPMD standard, and then used as input for a HiPACE++\,\cite{diederichs2022} simulation. The resulting propagation in the simulation is shown in Fig.~\ref{fig:experimental} (f) both in vacuum and in a plasma with a density of $2\times 10^{18}\,$cm$^{-3}$. A comparison with the equivalent Gaussian pulse is also shown, to illustrate the significant impact of using a realistic profile.

In this example a Gaussian temporal profile was used for simplicity, although experimentally measured temporal profiles can also be utilised in LASY. For this system, the laser pulse duration can be measured via self-referenced spectral interferometry using a commercial device (\emph{Fastlite Wizzler}) \cite{Oksenhendler2010}, although a variety of other techniques and commerical devices exist. The measurement provides the electric field amplitude and phase of the pulse envelope in either the spectral or temporal domain. LASY accepts data in both formats (indepent of measurement device) and so allows for this data to be readily added as a \emph{longitudinal profile}.

\section{From simulation to simulation}

\begin{figure}[h]
\begin{center}
\includegraphics[width=38pc]{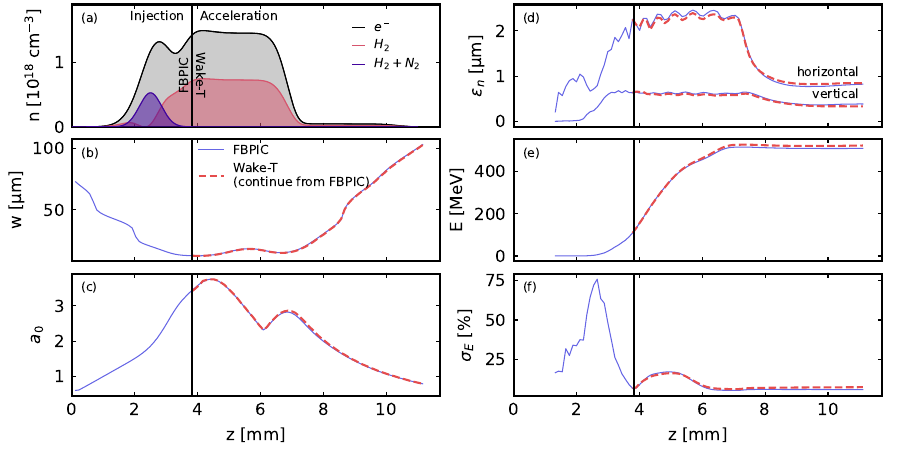}
\caption{Laser wakefield acceleration simulation using FBPIC for beam injection and Wake-T for the subsequent acceleration, using LASY to convert and transfer the laser profile between the two codes at $z=3.9$~mm. (a) plasma profile, see Ref.\,\cite{kirchen2021} for more details. (b) and (c) show the laser pulse width and peak normalized amplitude, respectively, comparing the workflow with both codes to a full FBPIC simulation. (d)-(f) show a similar comparison for the accelerated electron beam. Both laser and beam properties are accurately captured in the Wake-T simulation after being transferred from FBPIC.}
\end{center}\label{fig:sim1}
\end{figure}

Start-to-end workflows combine accuracy and computational efficiency by using the most appropriate code at every step of a physical process. An example is shown in Fig.\,\ref{fig:sim1} for a laser wakefield accelerator similar to that in Ref.\,\cite{kirchen2021}. The first part of the process is simulated with the electromagnetic PIC code FBPIC\,\cite{lehe2016}, to properly capture the injection of an electron beam in the wake of the laser near the downramp located at $z\sim3$~mm. After the end of the doped region ($H_2+N_2$), electron injection into the wake no longer occurs and the subsequent acceleration process can be efficiently modeled under the quasi-static approximation\,\cite{sprangle1990}. To make this transition, the laser pulse and the injected electron beam are transferred into a Wake-T\,\cite{pousa2019} simulation at $z\sim4$~mm.

\begin{figure}[h]
\begin{center}
\includegraphics[width=38pc]{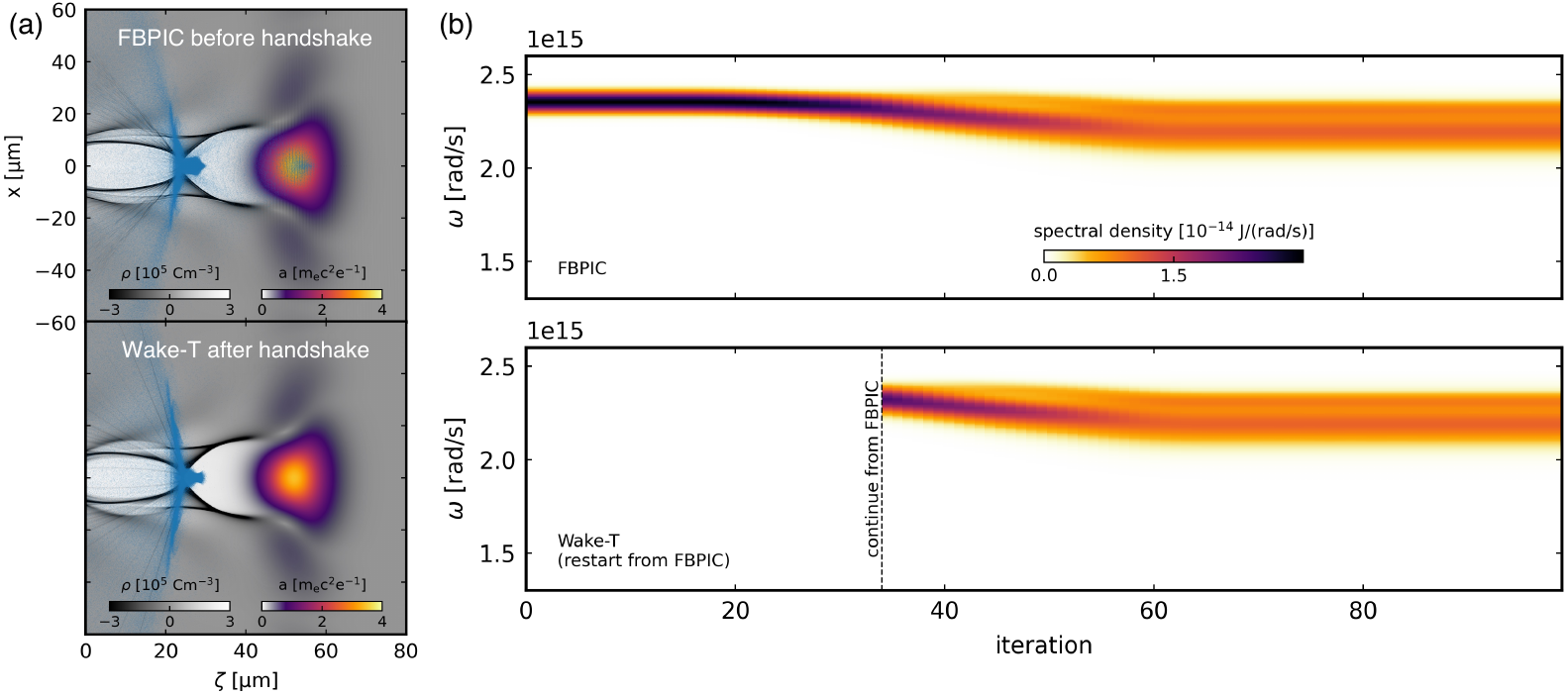}
\caption{(a) Snapshot of the wakefield process on the last time step with FBPIC (top) and on the first time step (bottom) with Wake-T. The laser amplitude is properly represented. (b) Spectrum of the laser pulse from the full FBPIC simulation (top) and from Wake-T after being transfered from FBPIC (bottom). The spectrum evolution is fully captured during the LASY operations.}
\end{center}\label{fig:sim2}
\end{figure}

LASY is used during this transition to convert the laser profile from a fully electromagnetic representation, including the fast oscillations at the carrier frequency, to an envelope model, as used in Wake-T. This conversion is performed in two steps. First, the envelope extraction is done using a 1D Hilbert transform along the longitudinal direction. Second, the laser field is converted from electric field to normalized vector potential $\boldsymbol{a}$, defined from the electric field as $\boldsymbol{E_{\perp}}=-\frac{m_ec}{e}\partial_t \boldsymbol{a}$. Importantly, both operations are performed preserving the full spectral, temporal and spatial profile of the laser pulse. This can be seen in Fig.\,\ref{fig:sim1} (b) and (c), where the laser propagation in the start-to-end workflow combining FBPIC and Wake-T is compared to the same simulation using only FBPIC for the whole propagation. Excellent agreement can be observed on the evolution of the main laser properties (pulse width and peak amplitude), as well as on the electron beam evolution (Fig.\,\ref{fig:sim1} d-f). More details about the transition between both codes can be seen in Fig.\,\ref{fig:sim2}. In particular, the laser spectrum is accurately preserved and continues to evolved as expected in the Wake-T simulation.

The full FBPIC simulation takes $\sim2$ hours running in parallel on 4 NVIDIA A100 GPUs, while the Wake-T section only takes $\sim7$ minutes on a single CPU core, so significant time and power gain can be achieved with such workflows, although a convergence study would be required for a quantitative comparison. Although this illustration presents a quasi-axisymmetric problem, LASY can equally support this workflow for more costly 3D simulations.

\section{Conclusion}

LASY is a versatile library to simplify the use of realistic profiles for higher-fidelity simulations and start-to-end workflows. It adopts community standards to encourage multi-code pipelines for higher efficiency. LASY is developed as part of an international collaboration, with the goal to combine efforts to expose more features (optics, phase manipulations, spatio-temporal coupling, propagators, filters, analytical profiles, readers, etc.) to the community in a user-friendly manner.

\section{References}
\bibliography{references}

\end{document}